\documentclass[preprint,12pt]{elsarticle}

\usepackage{subcaption}
\usepackage{amsmath}
\usepackage{mathtools}
\usepackage{autobreak}
\usepackage{nccmath}
\usepackage{latexsym}
\usepackage{amssymb}
\usepackage{amsfonts}
\usepackage{amsthm}
\usepackage{algorithm}
\usepackage[noend]{algpseudocode}

\usepackage{url}
\usepackage{floatflt}
\usepackage{diagbox}
\usepackage{caption}
\usepackage{subfig}

\usepackage{enumitem}
\algnewcommand\algorithmicforeach{\textbf{for each}}
\algdef{S}[FOR]{ForEach}[1]{\algorithmicforeach\ #1\ \algorithmicdo}
\algnewcommand{\LineComment}[1]{\State \(\triangleright\) #1}

\newcommand{\N}{\mathbb{N}} 
\newcommand{\J}{\mathcal{J}} 
\newcommand{\B}{\mathcal{B}} 
\newcommand{\Z}{\mathcal{Z}} 
\renewcommand{\S}{\mathcal{S}} 
\newcommand{\E}{\mathcal{E}} 
\newcommand{\comp}{\text{comp}}
\renewcommand{\deg}{\text{deg}}
\renewcommand{\time}{\text{time}}

\begin{document}

\begin{frontmatter}



\title{
  Computing Time-varying Network Reliability using Binary Decision Diagrams\tnoteref{fund}
}
\tnotetext[fund]{This work is partially supported by JSPS KAKENHI JP24K02931, JP22H03549, and JP22K17851.}


\author[1]{Yu Nakahata}
\ead{yu.nakahata@is.naist.jp}

\author[1]{Shun Arizono}
\ead{arizono.shun.ar6@is.naist.jp}

\author[1]{Shoji Kasahara}
\ead{kasahara@is.naist.jp}

\affiliation[1]{
  organization={
      Nara Institute of Science and Technology
  },
  country={Japan}
}



\begin{abstract}
Computing the reliability of a time-varying network, 
taking into account its dynamic nature, is crucial for networks that 
change over time, such as space networks, vehicular ad-hoc networks, and 
drone networks. 
These networks are modeled using temporal graphs, 
in which each edge is labeled with a time indicating its existence at a 
specific point in time. 
The time-varying network reliability is defined as the probability that a data 
packet from the source vertex can reach the terminal vertex, 
following links with increasing time labels (i.e., a journey), 
while taking into account the possibility of network link failures. 
Currently, the existing method for calculating this reliability involves 
explicitly enumerating all possible journeys between the source and terminal 
vertices and then calculating the reliability using the sum of disjoint 
products method. 
However, this method has high computational complexity. 
In contrast, there is an efficient algorithm that uses binary decision 
diagrams (BDDs) to evaluate the reliability of a network whose topology does 
not change over time. 
This paper presents an efficient exact algorithm that utilizes BDDs for computing the time-varying network reliability.
Experimental results show that the proposed method runs faster than the existing method up to four orders of magnitude.
\end{abstract}




\end{frontmatter}


\section{Introduction}\label{s:intro}
Networks such as optical networks and mobile networks are indispensable and 
ideally must be functional without failures. 
However, network links can fail or be disconnected due to events such as 
disasters. 
Thus, networks are designed taking possible failures of links into account.
Network reliability~\cite{perez-roses_sixty_2018} is an important 
quantitative measure to represent the robustness of networks against failures. 
Network reliability is defined as the probability that two specified nodes 
can communicate with each other given the fully functional nodes and the links 
that can fail independently with a certain probability. 
The network reliability is an important measure that represents 
not only the network robustness against failures, but also the 
fault-tolerance of computer architecture and the quality of services 
such as logistics and power transmission.

Network reliability has been studied for more than 60 years~\cite{perez-roses_sixty_2018, moore_reliable_1956}, and its computation is \#P-hard~\cite{valiant_complexity_1979, provan_complexity_1986},
which suggests it is theoretically difficult to compute efficiently.
Various methods have been proposed that attempt to compute 
the network reliability in efficient way 
including the ones~\cite{imai_computational_1999, hardy_k-terminal_2007}
that use binary decision diagrams (BDDs)~\cite{bryant_graph-based_1986}.
However, many of these methods consider a static network model where all links 
fail simultaneously, without taking into account the topological change in time.

Recently, space networks, vehicle ad-hoc networks, and drone networks 
are gaining popularity~\cite{alwis_survey_2021}.
These networks have the dynamic nature that the network topology changes over time
and are called time-varying networks (TVNs).
Computing the TVN reliability is crucial as their use cases include 
critical missions such as space missions, thus an efficient computation method is desired.
A TVN is modeled using a \emph{temporal graph} in which every edge has a time label of positive integer, and its reliability is defied as the probability that a data packet from the source node can reach the terminal node by following a path
composed of edges with non-decreasing time labels, i.e., a \emph{journey}~\cite{michail_introduction_2016}.
Chaturvedi et al.\ proposed an algorithm to compute the TVN reliability~\cite{chaturvedi_reliability_2018} by first enumerating paths explicitly, extracting journeys from them, and then
compute the reliability using the sum-of-disjoint products (SDP) method~\cite{chaturvedi_network_2016}.
However, this method has the high computational complexity that exponentially
increases with the number of vertices.
On the other hand, as mentioned earlier, there is an efficient algorithm for
the static network reliability that utilizes BDDs.

In this paper, we aim to improve the efficiency of TVN reliability
computation by extending the BDD-based method for computing network reliability from static networks to TVNs.
Especially, our main technical contribution is to propose an efficient method of constructing a ZDD~\cite{minato_zero-suppressed_1993} (a BDD-like data structure) representing a set of journeys in a given temporal graph.
By using a ZDD, we can compress the vast number of journeys in a compact way and expect the reduced computational cost.
The proposed method is an extension of the existing method to construct a ZDD representing a set of paths in a static graph, and thus we believe the algorithm for journeys is of independent interest.

The rest of this paper is organized as follows.
First, we briefly explain related work in Section~\ref{ss:related_work}.
In Section~\ref{s:preliminaries}, we introduce a temporal graph and a journey and give the formal definition of the TVN reliability. We also introduce BDDs/ZDDs and frontier-based search, which is a framework of algorithm to construct a BDD/ZDD in a top-down manner.
Next, we present our method for computing TVN reliability 
in Section~\ref{s:proposed_method}, and report experimental results in
Section~\ref{s:experiments}.
Finally, we describe concluding remarks in Section~\ref{s:conclusions}.

\subsection{Related Work}\label{ss:related_work}

Chaturvedi et al.\ proposed an algorithm to compute the TVN reliability~\cite{chaturvedi_reliability_2018}
by first enumerating all the paths from the source node to the terminal node, 
extracting only journeys, and then computing the reliability using the SDP
method~\cite{chaturvedi_reliability_2018}.
Temporal graphs used for the experiments were generated by assigning positive
integers from 1 to $T$ to each edge of a static complete graph with $n$
vertices, based on a certain probability (each edge may have multiple time labels, i.e., a sequence of non-decreasing time labels).
This method enumerates journeys by checking whether the Cartesian product of
sequences of time labels of a path is non-decreasing.
The time complexity of journey enumeration is $O(n!T^n)$, which increases
exponentially with the number of vertices.
Finally, the TVN reliability is computed using the SDP method.
The SDP method involves explicitly enumerating the events that include at least
one journey and the other events that are mutually exclusive to them, which 
suggests the high computational complexity.
Chaturvedi et al.\ reported the experimental results only for cases of 
$n=5$ and 6. 
In contrast, the static network reliability computation method that uses 
BDDs is able to compute network reliability with a few hundred links~\cite{imai_computational_1999,hardy_k-terminal_2007}.
In this paper, by extending the existing method that uses BDDs from static networks to TVNs, we aim to design an algorithm capable of computing the reliability of a TVN with hundreds of links.

\section{Preliminaries}\label{s:preliminaries}
In this section, we first describe a temporal graph and a journey,
and then define the TVN reliability.
We also describe a BDD/ZDD, a data structure used in our algorithm, and frontier-based search (FBS).

\subsection{Temporal Graphs and Journeys}\label{ss:temopral_graphs_and_journeys}
\begin{figure}[t]
    \centering
    \includegraphics{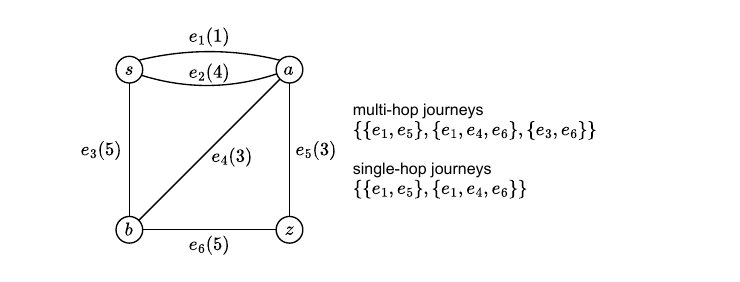}
    \caption{Example of a temporal graph. Numbers in parentheses indicate time labels.
    As examples, multi-hop and single-hop $s$-$z$ journeys are shown on the right side of the figure.}
    \label{fig:temporal_graph}
\end{figure}
A \emph{temporal graph} is an undirected graph (allowing multiple edges) that every edge has a 
\emph{time label} to indicate its existence at a specific point in time.
Let $G=(V, E, t)$ be a temporal graph where $V$ is a set of vertices,
$E=\{e_1,\dots,e_m\}$ is a set of undirected edges, and $t$ is 
a time label function $t\colon E\rightarrow\N$, where $\N$ is the set of positive integers.
Each edge has exactly one time label and exists only at the time of its value.
In a temporal graph, the set of vertices does not change over time,
but the set of edges does, which means 
connectivity of the vertices change over time.

Consider a sequence of edges $S=(e_{i_1},\dots,e_{i_k})$ that satisfies 
the following conditions:
\begin{itemize}
  \item $\forall j\in\{1,\dots,k-1\},
        e_{i_j}\textrm{ and }e_{i_{j+1}}~\textrm{share exactly one endpoint and }\\
        t(e_{i_j}) \le t(e_{i_{j+1}}), $ and
  \item $\forall j, j' \in \{1, \dots, k\}$ with $1\le j, j'\le k$ and $j' - j \ge 2$, $e_{i_j}$ and $e_{i_{j'}}$ do not share any endpoints.
\end{itemize}
For $S$, let $u$ be the endpoint of $e_{i_1}$ that is not shared with $e_{i_2}$ and 
let $v$ be the endpoint of $e_{i_k}$ that is not shared with $e_{i_{k-1}}$.
We say that $S$ is a \emph{$u$-$v$ journey} (or simply, a journey).
A consecutive subsequence of $S$ is also a journey, which we call a \emph{subjourney}.
When the time labels of edges in $S$ are non-decreasing, we say $S$ is \emph{multi-hop}, and when time labels are strictly increasing, we say $S$ is \emph{single-hop}.

If there exists a $u$-$v$ journey in $G$, we say $u$ is \emph{reachable} to $v$ and denote it as $u\rightsquigarrow_G v$.
$u\rightsquigarrow_G v\implies v\rightsquigarrow_G u$ is not always true.

An example of a temporal graph is shown in Figure~\ref{fig:temporal_graph}.
Temporal graphs are used to model networks in which the connections between nodes change over time, i.e., TVNs, because it is necessary to capture the time-varying characteristics of the connected components.

\subsection{Time-varying Network Reliability}\label{ss:tvn_rel}
A time-varying network is modeled using a temporal graph $G=(V,E,t)$, and
we regard a TVN as a temporal graph from now on.
Nodes in a TVN correspond to the vertices of $G$ and links correspond to 
edges of $G$.
We call the vertex from which a message is sent a \emph{source vertex} and 
denote it as $s$.
For the vertex at which a message arrives, we call it a \emph{terminal vertex} and denote it as $z$. 
Let $V[X]\coloneqq\bigcup_{\{u,v\}\in X}\{u,v\}$ where $X\subseteq E$.
The induced subgraph $G[X]\coloneqq (V[X],X,t')$ is also a temporal graph
where $t'\colon X\rightarrow\mathbb{N}$ is a time label function such that
$\forall e\in X,t'(e)=t(e)$.
If $s\rightsquigarrow_{G[X]}z$, $X$ is called a \emph{source-terminal reachable edge subset (STRES)}.

Let $\sigma(G)$ be the \emph{reliability} of a temporal graph $G$, and
$\S_G\subseteq 2^E$ be a family of all the STRESes of $G$.
Assuming $e_i\in E$ is independently deleted with the probability
$q(e_i)$ ($p(e_i)=1-q(e_i)$ is the survival rate),
the TVN reliability is defined as
\begin{equation}\label{eq:def_reliability}
  \sigma(G)\coloneqq\sum_{X\in \S_G}p(X),
\end{equation}
where
\begin{equation}\label{eq:content_of_reliability}
  p(X)\coloneqq\prod_{e_i\in X}p(e_i)\prod_{e_j\in E\setminus X}(1-p(e_j)).
\end{equation}

The TVN reliability evaluation problem is to compute the reliability $\sigma(G)$ given a temporal graph $G$ as an input graph.
If multi-hop journeys are allowed, we call the reliability \emph{multi-hop TVN reliability}.
If only single-hop journeys are allowed, we call the reliability \emph{single-hop TVN reliability}.
For the multi-hop TVN reliability, if all edges have the same time label, the problem is equivalent to the static network reliability evaluation.
Because the static network reliability evaluation is \#P-hard~\cite{valiant_complexity_1979,provan_complexity_1986}, the TVN reliability evaluation is also \#P-hard.
In contrast, for the single-hop TVN reliability, its evaluation is not a generalization of the static network reliability evaluation, thus we cannot say it is \#P-hard immediately.
However, to our best knowledge, no polynomial time algorithm to compute the single-hop TVN reliability is known.
Therefore, we avoid the explicit enumeration of STRESes, and by employing the implicit enumeration approach, we aim to construct an efficient algorithm.

\subsection{BDDs and ZDDs}\label{ss:bdd}
We use a BDD~\cite{bryant_graph-based_1986} for implicit enumeration of 
STRESes $\S_G$.
A \emph{BDD (binary decision diagram)} is a compact representation of Boolean functions, which is often used as an indicator function of a set family.
Here, we use each edge in $E$ as a variable such that for an edge subset
$X\subseteq E,e_i\notin X\ (e_j\in X)$ is assigned to $\mathtt{False}\ (\mathtt{True})$.

A BDD is a rooted directed acyclic graph $\B=(N,A)$ with a node set $N$ and an arc set $A$.
It has exactly one \emph{root node} $\rho$ and exactly two \emph{terminal nodes} $\bot$ and $\top$.\footnote{To avoid confusion, we use ``vertex'' and ``edge'' for temporal graphs and ``node'' and ``arc'' for BDD/ZDDs.}
Each non-terminal node $\alpha\in N$ has a \emph{label} $l(\alpha)\in\{1,\dots,m\}$,
and has exactly two outgoing arcs called the \emph{0-arc} and the \emph{1-arc}.
The node pointed by $x$-arc ($x\in\{0,1\}$) of $\alpha$ is called the \emph{$x$-child}, and denoted by $\alpha_x$.
$l(\alpha)<l(\alpha_x)$ holds if $\alpha_x$ is not a terminal node.

A directed path from $\rho$ to $\top$ represents a (possibly partial) variable 
assignment for which the represented Boolean function is $\mathtt{True}$.
If the path descends a 0-arc (1-arc) of a node $\alpha$, the variable
$e_{l(\alpha)}$ is assigned to $\mathtt{False}\ (\mathtt{True})$.

A BDD is obtained from a binary decision tree by applying the following rules as many times as possible.
\begin{enumerate}
  \item Delete $\alpha$ if $\alpha_0=\alpha_1$, and replace each arc ($\alpha', \alpha)$ by $(\alpha', \alpha_0\ (= \alpha_1))$.
  \item Share any two nodes $\beta,\beta'$ if $l(\beta)=l(\beta')$, 
        $\beta_0=\beta_0'$, and $\beta_1=\beta_1'$.
\end{enumerate}
These two rules eliminate the redundant nodes in the BDD, and
the obtained BDD is said to be \emph{reduced}.
Any BDD has the unique reduced form under the same variable order~\cite{bryant_graph-based_1986}.
Hereafter, a reduced BDD is simply called a BDD.

A BDD can represent a dense set family with less nodes~\cite{minato_zero-suppressed_2001}.
On the other hand, a \emph{zero-suppressed BDD (ZDD)}~\cite{minato_zero-suppressed_1993} is known to be a suitable data structure for a sparse set family.
A ZDD $\Z=(N,A)$ is also a rooted directed acyclic graph derived from a BDD.
A ZDD is obtained from a binary decision tree by applying the following two rules as many times as possible.
Note that only the first deletion rule is different from the one for a BDD.
\begin{enumerate}
  \item Delete $\alpha$ if $\alpha_1=\bot$ and replace each edge ($\alpha', \alpha)$ by $(\alpha', \alpha_0)$.
  \item Share any two nodes $\beta,\beta'$ where if $l(\beta)=l(\beta')$, 
        $\beta_0=\beta_0'$, and $\beta_1=\beta_1'$.
\end{enumerate}
Similarly to BDDs, the ZDD obtained by applying the above two rules is uniquely determined and said to be \emph{reduced}.
Hereafter, a reduced ZDD is simply called a ZDD.

\begin{figure}[t]
  \centering
  \begin{minipage}[b]{0.5\hsize}
    \centering
    \includegraphics[scale=0.9]{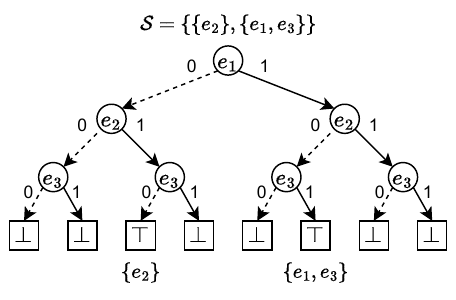}
    \subcaption{Binary decision tree.}\label{fig:binary_decision_tree}
  \end{minipage}
  \begin{minipage}[b]{0.2\hsize}
    \centering
    \includegraphics[scale=0.9]{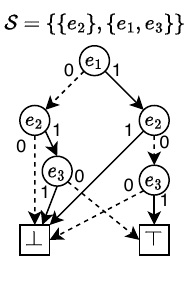}
    \subcaption{BDD.}\label{fig:bdd}
  \end{minipage}
  \begin{minipage}[b]{0.2\hsize}
    \centering
    \includegraphics[scale=0.9]{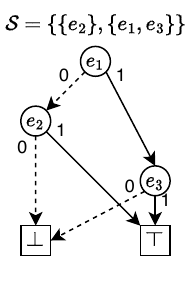}
    \subcaption{ZDD.}\label{fig:zdd}
  \end{minipage}
  \caption{A binary decision tree, a BDD and a ZDD. Solid arrows are 1-arcs and dashed arrows are 0-arcs.}\label{fig:trees}
\end{figure}

Figure~\ref{fig:trees} shows a binary decision tree, a BDD, and a ZDD representing the set family $\S=\{\{e_2\},\{e_1,e_3\}\}$.
Due to the sparsity of the set family, the ZDD has fewer nodes than the BDD.

\subsection{Frontier-Based Search}\label{ss:fbs}
To construct a BDD that represents the family $\S_G$ of source-terminal reachable edge subsets, we employ a framework called frontier-based search (FBS)~\cite{kawahara_frontier-based_2017}.
FBS is a general framework for the implicit enumeration of subgraphs that satisfy desired conditions.
In fact, we use FBS to construct a ZDD that represents the family of journeys in $G$.
Then, we obtain a BDD for $\mathcal{S}_G$ from the ZDD.
The details are discussed in Section~\ref{ss:overview}.
In this section, we describe the framework of FBS.

Given a function $C\colon 2^E\to\{0,1\}$, if $C(X)=1$ for an edge subset
$X\subseteq E$, then $X$ is said to \emph{have the property $C$}.
Let $P=\langle G,C\rangle$ be a problem to obtain all the possible edge
subsets of $E$ that have the property $C$.
The solution of $P$ is a family of edge subsets defined as
\begin{equation}\label{eq:def_solutions_with_the_characteristic}
  \E_P\coloneqq\{X\in 2^E : C(X)=1\}.
\end{equation}
Given a problem $P$, the algorithm constructs a BDD for $\E_P$ by processing edges individually as the exhaustive search;
the algorithm constructs the node set $N_i\coloneqq\{\alpha : l(\alpha) = e_i\}$ for $i=\{1,\dots,m\}$, and the arc set $A_x\coloneqq\{(\alpha,\alpha_x) : \alpha\in N_i\}$ for each $x\in\{0,1\}$.

The processed edges at the $i$-th step are denoted by 
$E^{\le i-1}\coloneqq\{e_1,\dots,e_{i-1}\}$, and
the unprocessed edges at the $i$-th step are denoted by 
$E^{\ge i}\coloneqq\{e_i,\dots,e_m\}$.
Let $\E(\alpha)\subseteq 2^{E^{\le i-1}}$ be a set of edge subsets corresponding
to the paths from the root to a node $\alpha\in N_i$.
Each node $\alpha\in N_i$ is associated with a subproblem of $P$ denoted by
$P_\alpha\coloneqq\langle G[E^{\ge i}],C_\alpha\rangle$ where the property 
$C_\alpha \colon 2^{E^{\ge i}} \to \{0, 1\}$ is defined as
\begin{equation}\label{eq:def_partial_solutions_with_the_characteristic}
  C_\alpha(X)=1~\iff~\forall Y\in\E(\alpha),C(X\cup Y)=1.
\end{equation}
For any pair of nodes $\beta,\beta'\in N_i$, $\beta$ and $\beta'$ are 
equivalent if $C_\beta(X)=1\iff C_{\beta'}(X)=1$ for any $X\in 2^{E^{\ge i}}$.
The algorithm merges some equivalent nodes into one node.

\begin{algorithm}[t]
  \caption{FrontierBasedSearch}\label{alg:fbs}
  \begin{algorithmic}
    \Require $E=\{e_1,\dots,e_m\}$
    \Ensure $\B$
    \State $N_1\gets\{\rho\},N_i\gets\emptyset$ for $i=2,\dots,m$
    \State Generate the terminals $\bot$ and $\top$
    \State $A_x\gets\emptyset$ for each $x\in\{0,1\}$
    \For{$i=1,\dots,m$}
      \For{$\alpha\in N_i$}
        \For{$x\in\{0,1\}$}
          \If{$\text{Prune}(\alpha,e_i,x)$}
            \State $A_x\gets A_x\cup\{(\alpha,\bot)\}$
          \ElsIf{$i=m$}
            \State $A_x\gets A_x\cup\{(\alpha,\top)\}$
          \Else
            \State $\beta\gets\mathrm{GenerateNode}(\alpha,e_i,x)$
            \If{$\exists\beta'\in N_{i+1},\phi(\beta)=\phi(\beta')$}
              \State $\beta\gets\beta'$
            \Else
              \State $N_{i+1}\gets N_{i+1}\cup\{\beta\}$
            \EndIf
            \State $A_x\gets A_x\cup\{(\alpha,\beta)\}$
          \EndIf
        \EndFor
      \EndFor
    \EndFor
    \State $N\gets(\bigcup_{i=1,\dots,m}N_i)\cup\{\bot,\top\},A\gets A_0\cup A_1$
    \State\Return $\B=(N,A)$
  \end{algorithmic}
\end{algorithm}

The framework of FBS is as follows.
Initially, the algorithm generates the node set $N_1=\{\rho\}$.
At the $i$-th step, the algorithm constructs $N_{i+1}$ using $N_i$ as follows.
For each node $\alpha\in N_i$, the algorithm generates its children;
$\E(\alpha_0)$ (resp.\ $\E(\alpha_1)$) is the set of the edge subsets that
$e_i$ is excluded from (resp.\ included in) the edge subsets of $\E(\alpha)$.
On generating a new child, the algorithm conducts the following procedures to 
reduce the number of nodes:
\begin{itemize}
  \item \emph{pruning}: Let Prune$(\alpha,e_i,x)$ be the function:
        \begin{equation}\label{eq:bot_prune}
          \text{Prune}(\alpha, e_i, x)=
            \begin{cases}
              \mathtt{True}& \E_{P_{\alpha_x}}=\emptyset,\\
              \mathtt{False}& \text{otherwise}.
            \end{cases}
        \end{equation}
        If Prune$(\alpha,e_i,x)$ returns $\mathtt{True}$,
        the $x$-child of $\alpha$ is $\bot$.
        Then the algorithm adds the $x$-arc $(\alpha,\bot)$ to $A_x$.
  \item \emph{merging}: Let $\beta$ be a child of $\alpha$.
        If $\beta$ and a node $\beta'\in N_{i+1}$ are equivalent,
        the algorithm sets $\beta'$ to $\beta$.
\end{itemize}
To apply these procedures efficiently, each node $\beta$ maintains an
additional information $\phi(\beta)$, referred to as a \emph{configuration} that
satisfies the condition that if $\phi(\beta)=\phi(\beta')$, 
$\beta$ and $\beta'$ are equivalent.
Note that the inverse is not required, which causes redundant node expansions.

Processing edges eventually leads to generating the node corresponding to $e_m$.
Let that node be $\alpha_m$.
If Prune$(\alpha_m, e_m, x)=\mathtt{False}$, the $x$-child of $\alpha_m$ is $\top$, and the algorithm adds $(\alpha_m,\top)$ to $A_x$.

Essentially, FBS is a dynamic programming using the configuration as
the state. It constructs a BDD as the structure derived from the table of
the dynamic programming.

The procedure of FBS is shown in Algorithm~\ref{alg:fbs}.
The function\linebreak
$\mathrm{GenerateNode}(\alpha,e_i,x)$ generates the $x$-child of $\alpha$.
The constructed BDD is not necessarily reduced because of possible redundant 
node expansions.
We apply the reduction rules until the BDD is reduced if needed.
The reduction can be performed in linear time in the number of nodes~\cite{knuth_art_2011}.

\section{Proposed Method}\label{s:proposed_method}
In this section, we present an algorithm for TVN reliability computation.

\subsection{The Overview of the Proposed Method}\label{ss:overview}
The overview of the proposed algorithm is as follows.
\begin{enumerate}[label=Step~\arabic*,leftmargin=15mm]
  \item Construct the ZDD that represents a journey family. We call the constructed ZDD \emph{a journey ZDD}.
  \item Construct the BDD that represents $\S_G$ from the journey ZDD.
  \item Compute the TVN reliability $\sigma(G)$ by a bottom-up dynamic
        programming on the BDD that represents $\S_G$.
\end{enumerate}
The proposed algorithm does not construct the BDD $\B$ for $\S_G$ directly.
Instead, it first constructs the ZDD for a journey family and converts it to $\B$.
This is because we predict the time complexity of the direct construction of 
$\B$ is greater than that of the journey ZDD construction.
Thus, we adopt the approach mentioned above.
The ZDD is used to represent a journey family because the family is 
expected to be sparse.
In contrast, because $\S_G$ is expected to be dense and we foresee 
that the BDD can represent $\S_G$ with less number of nodes,
the BDD is used for $\S_G$ representation.
To our best knowledge, an algorithm in Step~1 to construct a journey ZDD 
is yet known, and its design and implementation are our main technical contributions.
We describe the details of Step~1 in the next subsection.
In the latter of this subsection, we discuss how to implement Steps~2 and 3 using existing methods.

A family of STRESes is defined as
\begin{equation}\label{eq:def_sz_reachable_edge_sets}
  \S_G\coloneqq\{U\subseteq E : \exists J\in\J,J\subseteq U\},
\end{equation}
where $\J$ is a family of journeys.
We say $U$ is a \emph{superset} of $J$ if $J\subseteq U$.
The proposed algorithm constructs a BDD that represents $\S_G$ by performing
the superset operation~\cite{toda_superset_2015} on the journey ZDD.

Once the BDD $\B=(N, A)$ of $\S_G$ is obtained, the TVN reliability can be computed by applying dynamic programming to $\B$ in a bottom-up manner~\cite{imai_computational_1999, hardy_k-terminal_2007}.
Let each node $\alpha \in N$ hold the sum of probabilities $\psi(\alpha)$ of the subgraphs represented by the descendants of $\alpha$.
The probability values of $\bot$ and $\top$ are $\psi(\bot)=0$ and $\psi(\top)=1$, respectively.
The $\psi(\alpha)$ of a non-terminal node $\alpha \in N \setminus \{\bot, \top\}$ is calculated as
\begin{equation}\label{eq:compute_reliability_on_Bdd}
\psi(\alpha) = \psi(\alpha_1)p(e_{l(\alpha)}) + \psi(\alpha_0)(1 - p(e_{l(\alpha)})).
\end{equation}
From Equation~\eqref{eq:compute_reliability_on_Bdd}, $\sigma(G)$ equals $\psi(\rho)$ and the time required for their computation is linear to the number of nodes of $\B$.

\subsection{Frontier-Based Search for Journey Enumeration}\label{ss:fbsje}
In order to construct a journey ZDD efficiently, we propose the FBS for journey enumeration (FBSJE).
FBSJE is designed by the three main components of the FBS described in Section~\ref{ss:fbs}: configuration, pruning, and GenerateNode function.

Since the objective is to obtain the $s$-$z$ journeys, we define
\begin{equation}\label{eq:def_journey_characteristic}
C(X)=1\iff X \text{ is an $s$-$z$ journey of $G$.}
\end{equation}

\paragraph{Configuration}
Here we design a configuration for enumerating a family $\J$ of journeys.
We first describe the configurations needed to enumerate the journeys, and then we show that the configuration satisfies the conditions necessary to use the FBS.

For $i=1,\dots,m$, we define the \emph{$i$-th frontier} by
\begin{equation}\label{eq:def_frontier}
F_i\coloneqq V[E^{\le i-1}]\cap V[E^{\ge i}].
\end{equation}
Intuitively, a frontier is a set of vertices shared with both processed and unprocessed edges.

The existing FBS for path enumeration~\cite{kawahara_frontier-based_2017}
allows two pieces of information, the connected component and the degree, to be kept at the vertices of the frontier as a configuration.
In order to extend this to journey enumeration, we add a new piece of information to the frontier vertices as a part of configuration: the time label.
Specifically, for $\alpha\in N$, we define its configuration $\phi(\alpha)$ by a matrix whose columns are indexed by the frontier vertices, and each row represents the connected component (comp), the degree (deg), and the time label (time).
For readability, we divide the configuration into three arrays of connected components $\comp_\alpha$, degrees $\deg_\alpha$, and time labels $\time_\alpha$.
All the arrays are indexed by the frontier vertices.
A configuration with a subscript $\alpha$ means that it is a configuration of $\alpha\in N_i$.
The configuration has the following values.

\begin{equation}\label{eq:journey_config}
\comp_\alpha(v) =
\begin{cases}
0,&\parbox[t]{55mm}{if $v$ is connected to $s$} \\
1,&\parbox[t]{55mm}{if $v$ is connected to $z$} \\
c\in\{2,\dots,|F_i|+2\},&\parbox[t]{55mm}{if $v$ is connected to other vertex $v'$ than $s$ or $z$} \\
-1,&\text{if $v$ is an isolated point}
\end{cases}
\end{equation}

\begin{equation}
\deg_\alpha(v)\in\{0, 1, 2\}
\end{equation}

\begin{equation}
\time_\alpha(v) = 
\begin{cases}
t(e),&\parbox[t]{65mm}{if $v$ is an endpoint of a subjourney and $e\in E$ is incident to $v$}  \\
-1,&\text{otherwise}
\end{cases}
\end{equation}

For $\alpha \in N_i$ and $v \in F_i$, $\comp_\alpha(v)$ records the connected component of $v$ and $\deg_\alpha(v)$ records the degree of $v$.
For $\time_\alpha(v)$, we record the time label if and only if $v$ is an endpoint of a subjourney.

We assume that the FBSJE algorithm performs pruning in such a way as to ensure that the start and end points included in the frontier are either isolated points or end points of a journey, and that for other vertices, they are either isolated points, inner points of a journey, or end points of a journey.
If we can define a property that satisfies \eqref{eq:def_partial_solutions_with_the_characteristic} using the configuration,then we can define the conditions necessary for the configuration to use FBS, that is, it can be shown that if the configurations are the same, then regardless of how the processed edges have been adopted, the same way of adoption of unprocessed edges will produce a journey.
The following properties are defined to satisfy \eqref{eq:def_partial_solutions_with_the_characteristic}.

For $\alpha\in N_i$, we construct a temporal graph $G_\alpha=(V_\alpha, E_\alpha, t_\alpha)$ as follows.
Let $\hat{E}_\alpha$ be the set of edges incident to an internal vertex of a subjourney and $E_\alpha=E^{\ge i}\setminus \hat{E}_\alpha$.
Create a set of dummy vertices $I_\alpha$, each of which corresponds to a connected component of internal vertices of the journey that do not contain $s$ or $z$.
Let $i_j~(j=\comp_\alpha(v))$ be the dummy vertex in $I_\alpha$ and 
\begin{equation}
    V_\alpha\coloneqq \left( V[E^{\ge i}]\cup \{s,z\}\cup I_\alpha \right) \setminus\{v\in F_i : \deg_\alpha(v)=2\}.
\end{equation}
If there exists $v$ with $\comp_\alpha(v)=0$, that is, if there exists an endpoint $v$ of the subjourney containing $s$ in $F_i$, then let
\begin{equation}\label{eq:contracted_edge_s}
    e_\alpha^s \coloneqq\{s, v\},~t_\alpha(e_\alpha^s)=\time_\alpha(v)
\end{equation}
and add $e_\alpha^s$ to $E_\alpha$.
Similarly for the endpoint of the subjourney containing $z$, if there exists $v$ in $F_i$ satisfying $\comp_\alpha(v)=1$, then let
\begin{equation}\label{eq:contracted_edge_z}
    e_\alpha^z \coloneqq\{z, v\},~t_\alpha(e_\alpha^z)=\time_\alpha(v)
\end{equation}
and add $e_\alpha^z$ to $E_\alpha$.
For the frontier vertices other than $s$ or $z$, that is, the vertices of $\comp_\alpha(v)\ge 2$, we define
\begin{equation}
    e_\alpha^v \coloneqq\{v, i_j\},~t_\alpha(e_\alpha^v)=\time_\alpha(v)~(j=\comp_\alpha(v))
\end{equation}
and add $e_\alpha^v$ to $E_\alpha$.
The temporal graph $G_\alpha$ obtained by the above procedure is called a \emph{contracted graph} of $G$.
From the above assumption, $X\in\E_{P_\alpha}$ satisfies the condition that $X\cup E_\alpha$ is an $s$-$z$ journey of $G_\alpha$.
Therefore, we can define the property that satisfies \eqref{eq:def_partial_solutions_with_the_characteristic} as
\begin{equation}
C_\alpha(X) \iff X\cup E_\alpha \text{ is an $s$-$z$ journey of $G$}
\end{equation}
and it is shown that for any two nodes $\beta, \beta'\in N_i$, if $\phi(\beta)=\phi(\beta')$, then $\beta$ and $\beta'$ are equivalent.

\begin{figure}[t!]
  \centering
  \begin{minipage}[b]{0.33\hsize}
    \centering
    \includegraphics[width=\linewidth]{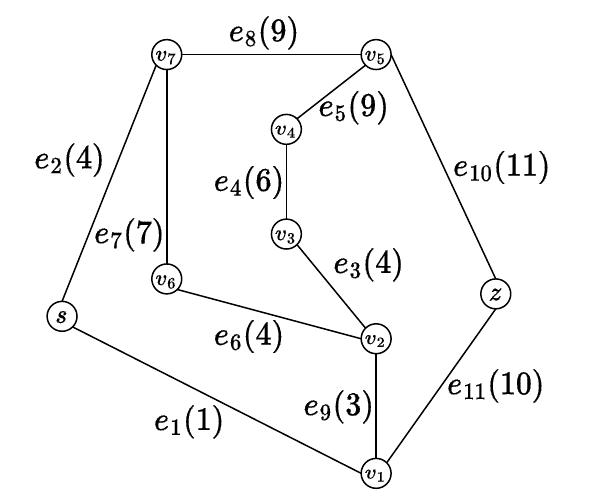}
    \subcaption{Temporal graph $G$.}\label{fig:fig:original_graph}
  \end{minipage}
  \begin{minipage}[b]{0.32\hsize}
    \centering
    \includegraphics[width=\linewidth]{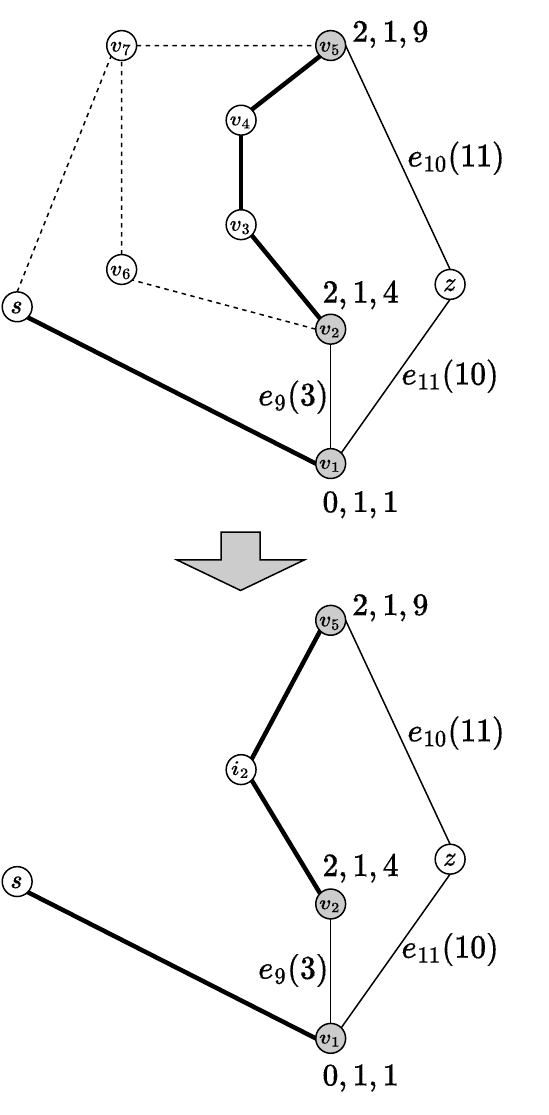}
    \subcaption{Contracted graph $G_\alpha$.}\label{fig:simplified_graph_a}
  \end{minipage}
  \begin{minipage}[b]{0.32\hsize}
    \centering
    \includegraphics[width=\linewidth]{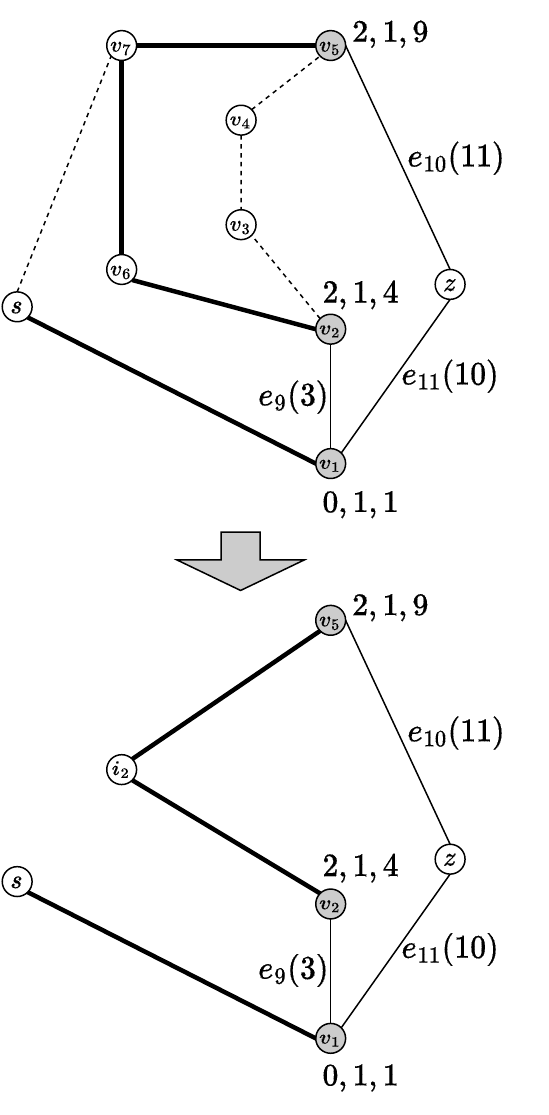}
    \subcaption{Contracted graph $G_{\alpha'}$.}\label{fig:simplified_graph_a_prime}
  \end{minipage}
  \caption{Temporal graph $G$ with two equivalent configurations of $\alpha,\alpha'\in N_8$. The thin solid edges are unprocessed edges, the thick solid edges are processed and adopted edges, and the dashed edges are processed but unadopted edges.
  The gray vertices $v_1,v_2,v_5$ are the frontier vertices.
  The $\comp, \deg, \time$ values of the frontier vertices are shown near the vertices like $x, y, z$.
  In both contracted graphs, it is common that the $s$-$z$ journey is completed when only $e_9$ and $e_{10}$ are adopted from the unprocessed edges.
  Although the adopted edges are different between $\alpha$ and $\alpha'$, the configuration values are the same, so the ways of adopting unprocessed edges to be a solution are the same.
  The decision of whether an edge subset is a journey or not can be made on a contracted graph defined only from configurations.}\label{fig:equal_configs}
\end{figure}

Figure~\ref{fig:equal_configs} shows a temporal graph with two equivalent configurations.
When two configurations are equivalent, the corresponding nodes are shared according to the node sharing rule of ZDDs.

\paragraph{$\mathbf{Prune}(\alpha, e_i, x)$}
We design the pruning procedure to be consistent with the aforementioned assumptions.
Given a ZDD node $\alpha$, an edge $e_i$ to process, and a variable $x\in\{0,1\}$ indicating whether the edge is adopted or not, if it is determined that there are no $s$-$z$ journeys in the subproblem $P_\alpha$, then $\mathrm{Prune}(\alpha, e_i, x)$ returns $\mathtt{True}$.
The following are the sufficient conditions that there are no $s$-$z$ journeys in $P_\alpha$.
\begin{enumerate}
\item Degree of $s$ or $z$ exceeds 2.
\item Degree of a vertex other than $s$ or $z$ exceeds 3. 
\item Degree of $s$ or $z$ is 0 and all edges incident to the vertex have been processed.
\item $s$ and $z$ are disconnected and no longer will be connected.
\item A cycle is created.
\item A sequence of time labels in a subjourney violates the definition of a journey.
\end{enumerate}
We call the above conditions \emph{(pruning) conditions} and refer them by the numbers (e.g., condition 1 means the first one.)

\begin{figure}[t!]
\centering
\includegraphics{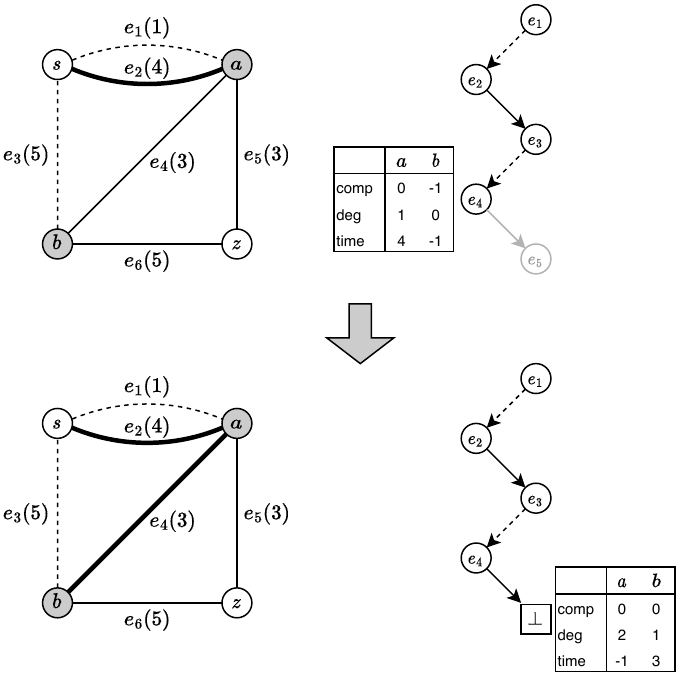}
\caption{An example of pruning by time labels.
In the graphs on the left, thin solid edges are unprocessed edges, thick solid edges are processed and adopted edges, and dashed edges are not processed but unadopted edges. The gray vertices $a, b$ are the frontier vertices, and the node with label $e_4$ is $\alpha$.
The example shows the situation when $e_4$ is about to be adopted. 
The upper part of the figure shows the temporal graph and the corresponding ZDD before updating the configuration, and the lower part shows those after the update.
Since the time label $t(e_4)$ of the endpoint $a$ of the journey starting from $s$ is smaller than $\time_\alpha(a)$, it is determined that it will not be a journey if $e_4$ is adopted, and therefore $\mathrm{Prune}(\alpha, e_4, 1)$ returns $\mathtt{True}$ by the condition 6.}
\label{fig:prune}
\end{figure}

Check the connected components, degree, and time labels of the vertices of the frontier, and if any one of the above conditions is true, execute the pruning, i.e., $\mathrm{Prune}(\alpha, e_i, x)$ returns $\mathtt{True}$.
Figure~\ref{fig:prune} shows an example of pruning based on the condition 6.

\begin{algorithm}[t]
\caption{Prune}\label{alg:Prune}
\begin{algorithmic}
\Require $\alpha, e_i=\{u, v\}, x$
\Ensure $\mathtt{True}$ or $\mathtt{False}$
\If{$x=1$}
    \If{$\comp_\alpha(u) = \comp_\alpha(v)$} \Comment{A cycle is generated (condition 5)}
        \State\Return $\mathtt{True}$
    \EndIf
    \If{CheckTimeCondition$(\alpha, e_i)=\mathtt{False}$}
        \State\Return $\mathtt{True}$ \Comment{condition 6}
    \EndIf
\EndIf
\State Copy $\alpha$ to $\alpha'$
\State $\text{UpdateInfo}(\alpha', e_i, x)$
\ForEach{$w\in \{u, v\}$}
    \If{$(w=s \lor w=z) \land (\deg_{\alpha'}(w) > 1)$}
        \State\Return $\mathtt{True}$ \Comment{condition 1}
    \ElsIf{$(w\neq s \land w\neq z) \land (\deg_{\alpha'}(w) > 2)$}
        \State\Return $\mathtt{True}$ \Comment{condition 2}
    \EndIf
\EndFor
\ForEach{$w\in F_{i+1}\setminus F_i$} \Comment{For all vertices leaving the frontier}
    \If{$(w=s \lor w=z) \land (\deg_{\alpha'}(w) \neq 1)$}
        \State\Return $\mathtt{True}$ \Comment{condition 3}
    \ElsIf{$(w\neq s \land w\neq z) \land (\deg_{\alpha'}(w)\neq 0) \land (\deg_{\alpha'}(w)\neq 2)$} 
        \State\Return $\mathtt{True}$ \Comment{condition 4}
    \EndIf
\EndFor
\State\Return $\mathtt{False}$
\end{algorithmic}
\end{algorithm}

Algorithm~\ref{alg:Prune} shows the function $\mathrm{Prune}(\alpha, e_i, x)$.
The function CheckTimeCondition, when $e_i=\{u, v\}$ is adopted, checks the relation between the union set $J_{\mathrm{new}}=J_u\cup e_i\cup J_v$ of journeys $J_u$ and $J_v$ with $u, v$ as the endpoints, and $\time_\alpha(u), \time_\alpha(v), t(e_i)$.
If $J_{new}$ does not satisfy the condition of being a journey, the function returns $\mathtt{False}$, otherwise it returns $\mathtt{True}$.
Whether a journey is multi-hop or single-hop is determined inside the CheckTimeCondition function by the equality sign is added to the inequality sign that indicates the relationship between the two time labels.
When checking the equality of time labels, the function checks whether there is a vertex on the frontier that corresponds to $u$, and if so, it checks whether that vertex has a time label.
If so, it is necessary to check the time label of that vertex as well.
The same is true for $v$, and since the computational complexity of this procedure is $O(|F_i|)$, the function CheckTimeCondition has a complexity of $O(|F_i|)$.
The other conditions in $\mathrm{Prune}$ can be checked in $O(1)$ time, and thus the time complexity of $\mathrm{Prune}$ is $O(|F_i|)$.

The function UpdateInfo takes a configuration as a reference and updates the connected components, degree, and time labels, as shown in Algorithm~\ref{alg:UpdateInfo}.
The function UpdateInfo merges the connected components of $u$ and $v$ to the smaller of $\comp_\alpha(u)$ and $\comp_\alpha(v)$ only when $e_i=\{u, v\}$ is adopted, and update the degree as $\deg_\alpha(u)\gets\deg_\alpha(u )+1, \deg_\alpha(v)\gets\deg_\alpha(v)+1$.
If the degree after update is 1, then update time labels as $\time_\alpha(u)\gets t(e_i), \time_\alpha(v)\gets t(e_i)$.
If the degree after update is 2, let $\time_\alpha(u)\gets -1, \time_\alpha(v)\gets -1$.
Since it takes $O(|F_i|)$ to update the connected components and $O(1)$ to update the degrees and time labels, the computational complexity of UpdateInfo function is $O(|F_i|)$.

\begin{algorithm}[t]
\caption{UpdateInfo}\label{alg:UpdateInfo}
\begin{algorithmic}
\Require $\alpha, e_i, x$
\State $e_i=\{u, v\}$
\If{$x=1$}
\If{$\comp_\alpha(u) = \comp_\alpha(v) = -1$}\Comment{new connected component}
    \State Let $c \ge 2$ be the minimum integer absent in $\comp_\alpha$
    \State $\comp_\alpha(u) \gets c$
    \State $\comp_\alpha(v) \gets c$
\Else
    \State $c_\text{max}\gets \max\{\comp_\alpha(u), \comp_\alpha(v)\}$
    \State $c_\text{min}\gets \min\{\comp_\alpha(u), \comp_\alpha(v)\}$
    \ForEach{$w\in F_i$} \Comment{merge connected components}
        \If{$\comp_\alpha(w)=c_\text{max}$}
            \State $\comp_\alpha(w)\gets c_\text{min}$
        \EndIf
    \EndFor
\EndIf
\ForEach{$w\in \{u,v\}$}
    \State $\deg_\alpha(w)\gets\deg_\alpha(w)+1$ \Comment{increment the degree of $w$}
    \If{$\deg_\alpha(w)=1$} 
        \State $\time_\alpha(w)\gets t(e_i)$ \Comment{if the degree after update is 1, record $t(e_i)$}
    \ElsIf{$\deg_\alpha(w)=2$} 
        \State $\time_\alpha(w)\gets -1$ \Comment{if the degree after update is 2, record $-1$}
    \EndIf
\EndFor
\EndIf
\end{algorithmic}
\end{algorithm}

\paragraph{$\mathbf{GenerateNode}(\alpha, e_i, x)$}
The primary role of the $\mathrm{GenerateNode}$ function is to update the configuration.
Since the procedure for updating the configuration is similar to that of the $\mathrm{UpdateInfo}$ function, we explain the additional part of $\mathrm{GenerateNode}$ against $\mathrm{UpdateInfo}$.
The difference from the $\mathrm{UpdateInfo}$ function is that it prepares nodes with label $e_{i+1}$.
First, a node $\beta$ with label $e_{i+1}$ is created and $\phi(\alpha)$ is copied to $\phi(\beta)$.
Next, the configuration is updated as follows.
\begin{enumerate}
\item If $x=0$, do not update the configuration. If $x=1$, execute\linebreak $\mathrm{UpdateInfo}(\beta, e_i, 1)$.
\item Remove the column corresponding to the vertex of $F_i\setminus F_{i+1}$ that will be removed from the frontier, and
    insert and initialize the columns corresponding to the vertices of $F_{i+1}\setminus F_i$ that are newly added to the frontier. For the configuration of the new frontier vertex $v\in F_{i+1}\setminus F_i$, we set $\comp_\beta(v)\gets -1, \deg_\beta(v)\gets 0, \time_\beta(v)\gets -1$.
\end{enumerate}
Since $|F_i\setminus F_{i+1}|$ and $|F_{i+1}\setminus F_{i}|$ are at most 2, the number of columns to be deleted or added is at most a constant.
Since the computational complexity of the node copy is $O(|F_i|)$ and that of the UpdateInfo function is $O(|F_i|)$, the complexity of the GenerateNode function is $O(|F_i|)$.

\subsection{Computational Complexity of FBSJE}\label{ss:complexity_of_fbsje}
In this subsection, we discuss the computational complexity of FBSJE.
For this purpose, we consider $|N_i|$, the number of nodes with label $e_i$.
$|N_i|$ is at most the number of possible patterns of the matrix $\phi$ for the nodes of label $e_i$.
The degree of a vertex $v$ in $F_i$ is 0, 1, or 2.
Since the number of connected components is at most $|F_i|$, the number of time labels is at most $T$, and the number of frontier vertices is $|F_i|$, then
$|N_i|$ is at most $(3 \cdot |F_i| \cdot T)^{|F_i|}$.
Since the computational complexity of $\mathrm{Prune}$ is $O(|F_i|)$,
and the computational complexity of GenerateNode is $O(|F_i|)$,
the computational complexity of FBSJE is
\begin{equation}
O\left(\sum_{i=1}^m |F_i|\cdot (3 \cdot |F_i| \cdot T)^{|F_i|}\right).
\end{equation}
The merit of FBSJE is that it is not directly affected by the size of the input temporal graph.
This suggests that journey enumeration can be applied to larger temporal graphs than the existing explicit enumeration algorithms.
The computational complexity of FBSJE increases exponentially with frontier size.
Since the frontier size depends on the order of edges to be processed, ordering the edges in such a way that the maximum frontier size is minimized can contribute to reducing the computation time.
The problem of finding the edge order that minimizes the maximum frontier size is NP-hard, and thus a heuristic method has been proposed~\cite{inoue_acceleration_2016}.


\section{Experiments}\label{s:experiments}
Computer experiments were conducted to verify the effectiveness of the proposed method.
The proposed method and the part of journey enumeration of the existing method were implemented in C++ (g++8.5.0 with the -O3 option).
The proposed method was implemented using TdZdd library, a highly optimized implementation of FBS.\footnote{\url{https://github.com/kunisura/TdZdd}}
The reliability calculation part of the existing method was implemented using the SDP method published by Chaturvedi et al.\footnote{\url{https://www.scrivenerpublishing.com/MatlabPrograms.rar}}
Note that only the program for the SDP method is implemented in MATLAB, which has the limitation that it can only handle instances with 53 or fewer edges.
As for the algorithm for the superset operation, the library used includes one that outputs a ZDD.
We decided to compare the computation time with that of our original implementation that outputs a BDD.
For convenience, we refer to the proposed method that outputs BDDs after the superset operation as Method~B, and to the method that outputs ZDDs as Method~Z.
We used a machine with Intel Xeon Gold 6238L CPU (22 cores $\times$ 4), 512 GB memory (allocated by slurm job scheduler), and RedHat Enterprise Linux 8.5 OS.

\subsection{Datasets}\label{ss:datasets}
The temporal graph instances used in the experiment were created by assigning a time label with a certain probability to each edge of the static graph, based on Chaturvedi et al.~\cite{chaturvedi_reliability_2018} 
The time label assigned to each edge of a static graph is an integer between 1 and $T$.
Assigning a time label to an edge of a static graph generates an edge of a temporal graph. Here, the probability of assigning a time label to an edge of a static graph is uniformly set to 0.5.
Since the assignment of time labels is done probabilistically, there are cases where no time labels are assigned at all.
Therefore, instances of the temporal graph are created iteratively until the edge containing the starting or ending point is included in the temporal graph.
The survival probability of each edge is uniformly set to 0.9 for TVN reliability evaluation.


The static graphs considered were complete graphs and grid graphs of height 3.
For both types of graphs, a breadth-first edge order was used such that the maximum frontier size is minimized.
Complete graphs were prepared with $n \in \{3, 4, \dots, 10\}$ vertices and $T=n-1$.
Grid graphs of height 3 were prepared with width $w \in \{3, 4, \dots, 10\}$ and $T=2w$.
For a complete graph with $n$ vertices, let $V=\{v_1,\dots,v_n\}$ be the vertex set, where $s = v_1$ and $z = v_n$.
For a grid graph of height $h$ and width $w$, when the vertex set is $V=\{v_1,\dots,v_{hw}\}$, we set $s = v_1$ and $z = v_{hw}$ so that $s$ and $z$ are on the diagonal of the grid graph.
To measure the average computation time, 100 instances of temporal graphs of the complete graph were prepared for each $n$.
Similarly, 100 instances of the grid graph were prepared for each $w$.
The program is executed with 100 instances for each $n$ as input in the case of a complete graph.
If the computation time of the program with the $i$-th instance of $n$ exceeds 2 hours, the program execution is stopped, and the experiment on the $j\ (\ge i)$-th instances of $n$ and all instances of $n'\ (> n)$ was terminated.
Similarly for grid graphs, when the computation time of the program with the $i$-th instance of $w$ exceeded 2 hours, the experiment for the $j\ (\ge i)$-th instances of $w$ and all instances of $w'\ (> w)$ was terminated.

\subsection{Comparison with the Existing Method}\label{ss:vs_extg}
\begin{table}[t]
  \centering
  \caption{Comparison of the proposed method and the existing method for complete graphs.
  Let $m_\text{avg.}$ be the average number of edges, $t_\text{extg.}$ be the average computation time of the existing method, $t_\text{B}$ be the average computation time of Method~B, and $t_\text{Z}$ be the average computation time of Method~Z.}\label{tab:vs_extg}
\subcaption{Results for multi-hop reliability.}
\begin{tabular}{rrrrrrr}\hline
    $n$ & $m_\text{avg.}$ & $t_\text{extg.}~\text{(s)}$ & $t_\text{B}~\text{(s)}$ & $t_\text{extg.}/t_\text{B}$ & $t_\text{Z}~\text{(s)}$ & $t_\text{extg.}/t_\text{Z}$ \\ \hline
 3 &     3.0 & 1.13e-01 & 4.49e-04 &   252 & 4.85e-04 &  233 \\
 4 &     9.0 & 3.16e-01 & 4.96e-04 &   637 & 5.45e-04 &  580 \\
 5 &    20.0 & 7.45e-01 & 6.75e-04 &  1105 & 8.80e-04 &  847 \\
 6 &  37.5 & 1.50e+02 & 4.02e-03 & 37372 & 1.81e-02 & 8309 \\
\end{tabular}
\vspace{3mm}

\subcaption{Results for single-hop reliability.}
\begin{tabular}{rrrrrrr}\hline
$n$ & $m_\text{avg.}$ & $t_\text{extg.}~\text{(s)}$ & $t_\text{B}~\text{(s)}$ & $t_\text{extg.}/t_\text{B}$ & $t_\text{Z}~\text{(s)}$ & $t_\text{extg.}/ t_\text{Z}$ \\ \hline
 3 &     3.0 & 1.30e-01 & 4.64e-04 & 279 & 4.93e-04 & 263 \\
 4 &     9.0 & 4.11e-01 & 5.04e-04 & 814 & 5.31e-04 & 774 \\
 5 &    20.0 & 3.32e-01 & 6.11e-04 & 544 & 6.69e-04 & 497 \\
 6 &  37.5 & 1.02e+00 & 1.28e-03 & 800 & 1.99e-03 & 514 \\
\end{tabular}
\end{table}
In this subsection, we show the comparison between the proposed method (Section~\ref{s:proposed_method}) and the existing method~\cite{chaturvedi_reliability_2018}.
The comparison is conducted for complete graphs with $n \in \{3, 4, 5, 6\}$ vertices.
Table~\ref{tab:vs_extg} shows the results.
As for column labels, $n$ represents the number of vertices in the complete graph and the $m_\text{avg.}$ represents the average number of edges.
Columns $t_\text{extg.}, t_\text{B}$, and $t_\text{Z}$ denote the average computation time of the existing method, Method~B, and Method~Z, respectively.
The $t_\text{extg.}/t_\text{B}$ column indicates the computation time of the existing method divided by the computation time of Method~B. It means how much faster Method~B is compared to the existing method.
The $t_\text{extg.}/t_\text{Z}$ column similarly indicates how many times faster Method~Z is compared to the existing method.

For all $n$, the average computation time of the proposed methods are smaller than those of the existing method.
In the multi-hop case, Method~B is up to about 37,372 times faster and Method~Z up to about 8,309 times faster.
In the single-hop case, Method~B is up to 814 times faster and Method~Z up to 774 times faster.
Therefore, the proposed method can compute the TVN reliability more efficiently than the existing method.
Method~B tends to be faster than Method~Z.
The reason for this is that the computational time for the superset operation and the reliability calculation of Method~B is smaller than that of Method~Z, which will be explained in detail in the following subsection.

\subsection{Analysis of the Proposed Method}\label{ss:vs_proposed}
\begin{figure}[t]
\centering
\includegraphics[height=0.8\textheight]{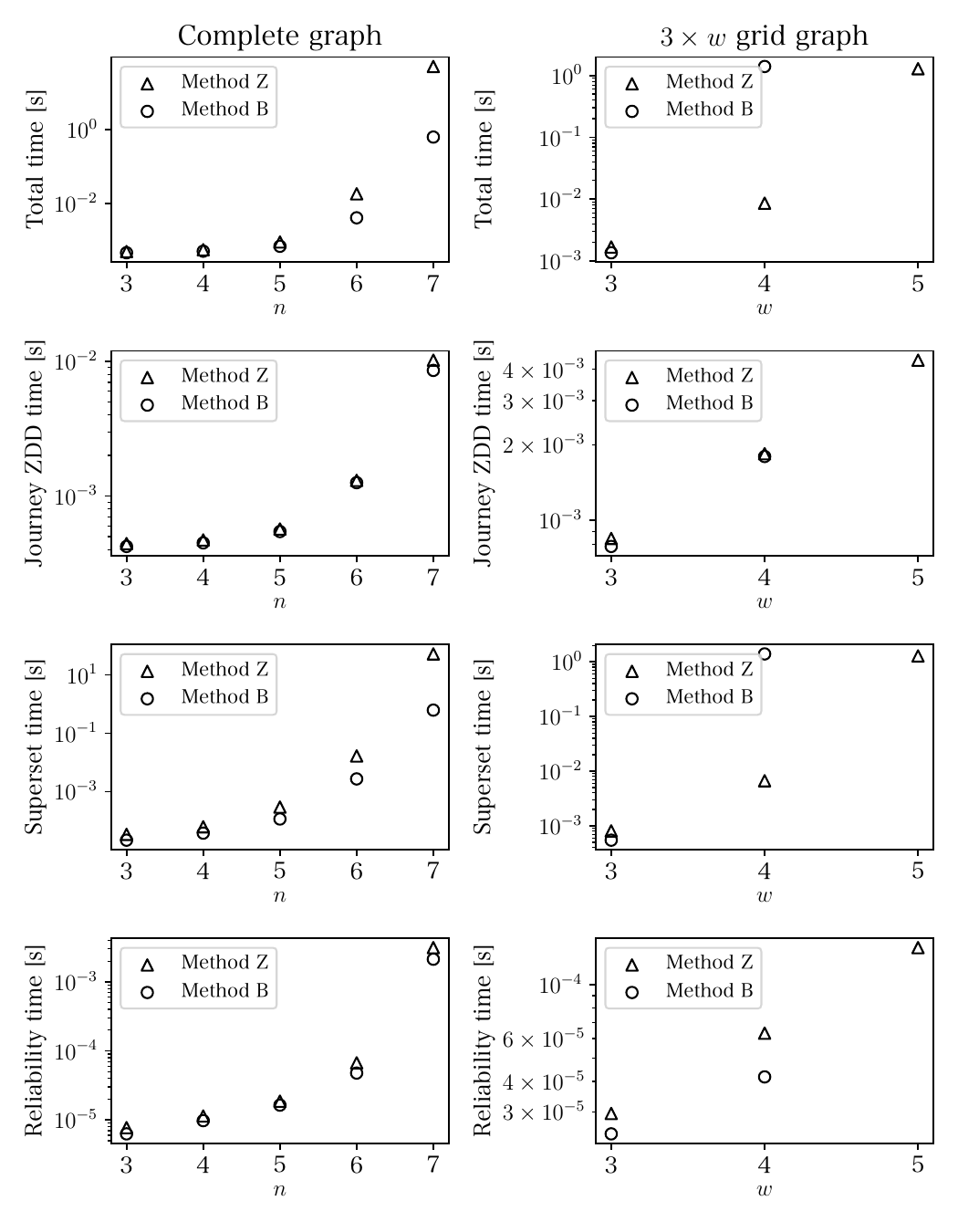}
\caption{Comparison of computation time for Methods~B and Z for the multi-hop reliability. Missing points indicate the time out of 2 hours.}
\label{fig:vs_proposed_m}
\end{figure}

\begin{figure}[t]
\centering
\includegraphics[height=0.8\textheight]{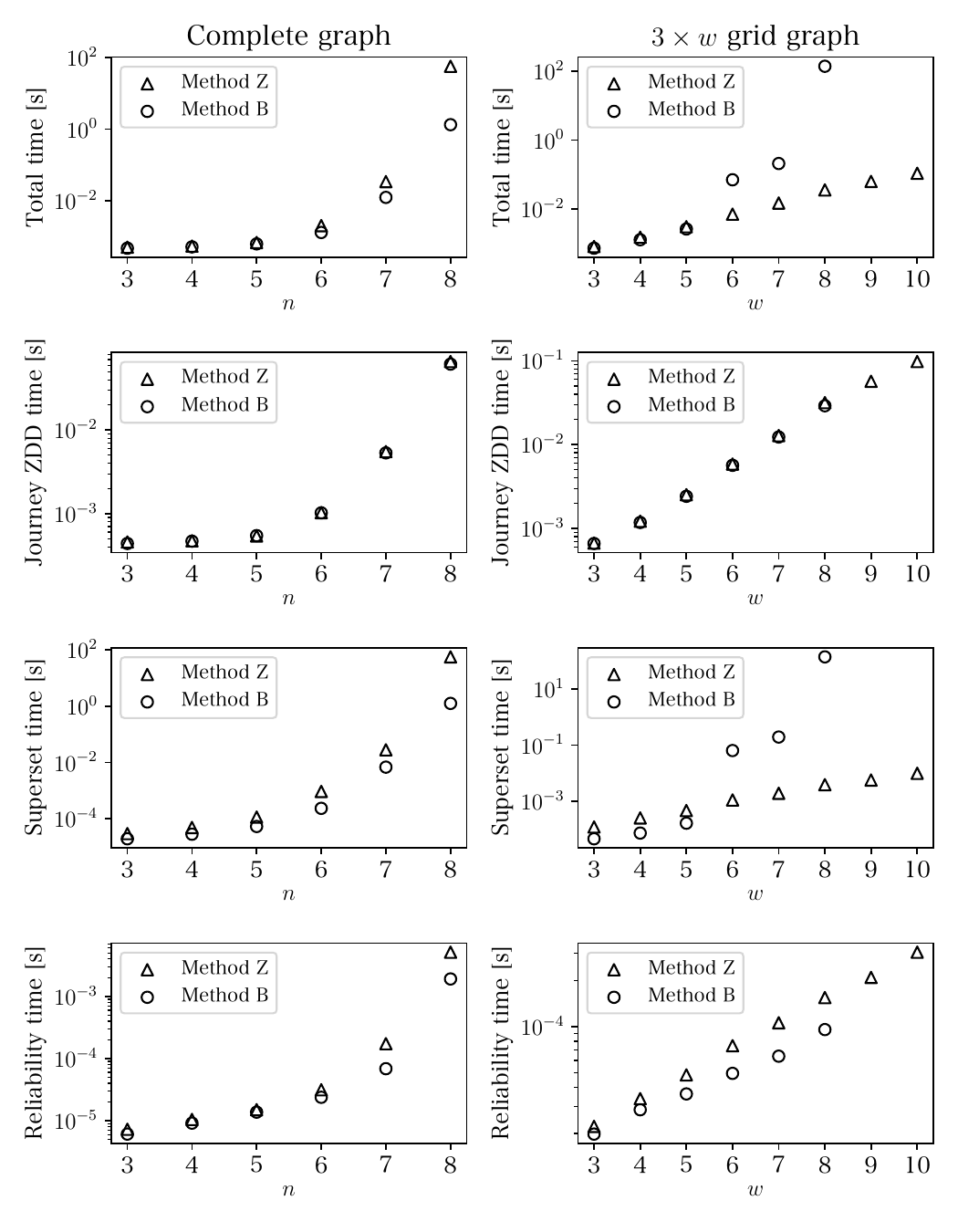}
\caption{Comparison of computation time for Methods~B and Z for the single-hop reliability. Missing points indicate that the time out of 2 hours.}
\label{fig:vs_proposed_s}
\end{figure}

The proposed method consists of three steps as stated in Section~\ref{ss:overview}.
In order to evaluate not only the overall computation time, but also the computation time for each step, scatter plots of the computation time for Methods~B and Z are shown in Figures~\ref{fig:vs_proposed_m} and \ref{fig:vs_proposed_s}, respectively.
In each figure, the scatter plots in the left column are for the complete graphs, with computation time on the vertical axis and the number of vertices $n$ on the horizontal axis.
The scatter plots in the right column are for grid graphs, with computation time on the vertical axis and grid graph width $w$ on the horizontal axis.
The scatter plots in the first row show the overall computation time, and the plots in the second row show the computation time for the construction of the journey ZDD.
The third row shows the computation time for the superset operation, and the last row shows the computation time for reliability by dynamic programming.

\begin{figure}[t]
\centering
\begin{minipage}[t]{\hsize}
\includegraphics[width=0.99\textwidth]{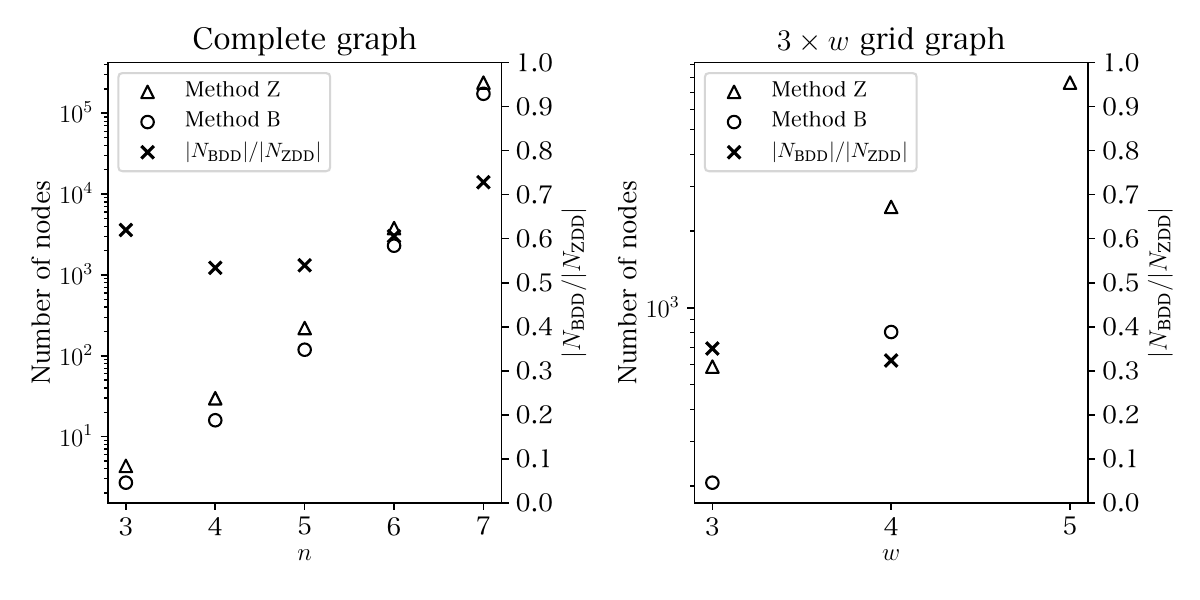}
\subcaption{Results for the single-hop reliability.}
\label{fig:n_nodes_m}
\end{minipage}
\begin{minipage}[t]{\hsize}
\includegraphics[width=0.99\textwidth]{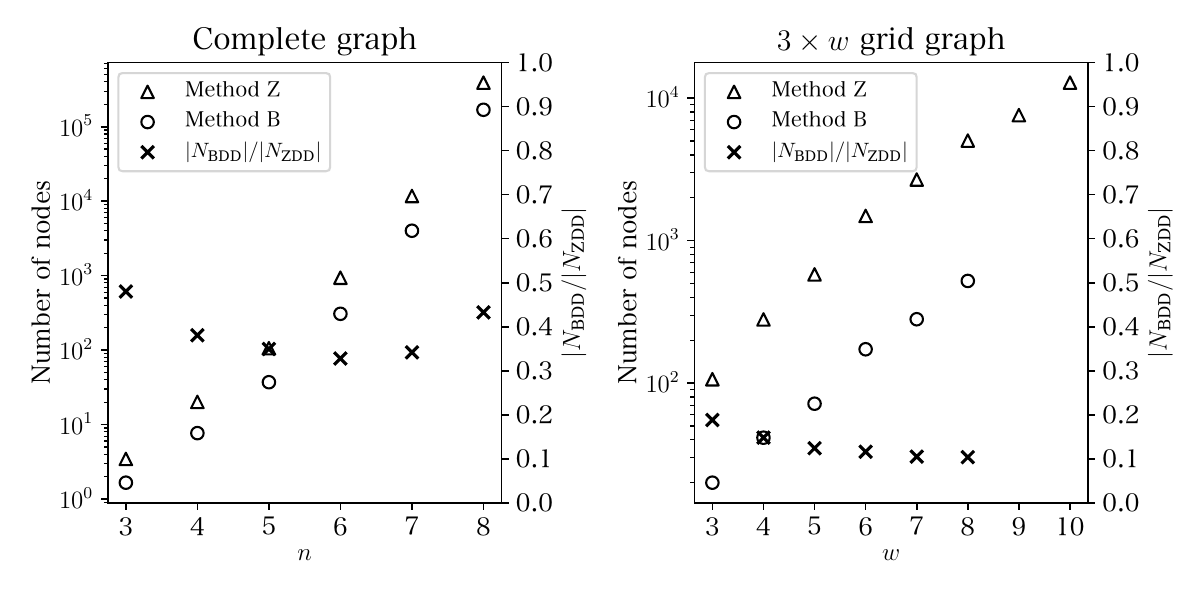}
\subcaption{Results for the single-hop reliability.}
\label{fig:n_nodes_s}
\end{minipage}
\caption{Comparison of the memory efficiency of Methods~B and Z. $N_\text{BDD}$ and $N_\text{ZDD}$ represents the numbers of BDDs and ZDDs, respectively.}
\label{fig:n_nodes}
\end{figure}

Method~B is faster than Method~Z for complete graphs, but Method~Z tends to be faster for grid graphs.
This can be attributed to the difference in computational time for the superset operation.
For complete graphs, Method~B's superset and reliability computation times are smaller than those of Method~Z, which may explain why Method~B is faster.
On the other hand, in the case of grid graphs, the computation time for the reliability computation of Method~B is smaller than that of Method~Z, but the computation time for the superset operation is larger than that of Method~Z. Unlike in the case of complete graphs, Method~Z is considered to be faster.
There are two possible reasons for the difference in superset computation time between complete graphs and grid graphs: the effect of edge density and the effect of the implementation method, but the latter seems to be the main cause.
The superset operation in Method~Z is highly optimized because it uses functions from a library that was prepared in advance, but the superset operation in Method~B was implemented independently based on the existing method~\cite{toda_superset_2015}, which is thought to be the reason for this trend.
The average time taken to construct the journey ZDD and the average computation time for the reliability calculation by dynamic programming is less than 0.1 second, but the average time taken for the superset operation is approximately 57 seconds at the maximum, suggesting that the superset operation is the bottleneck.

From Figures~\ref{fig:vs_proposed_m} and \ref{fig:vs_proposed_s}, we can see that there is a difference in the computation time for the reliability calculation depending on the methods.
To explain the reason for this, Figure~\ref{fig:n_nodes} shows scatter plots of the numbers of nodes in BDDs and ZDDs representing STRESes for Methods~B and Z, respectively.
Since the computational complexity of the reliability calculation is linearly proportional to the number of nodes, the computational time can be reduced by reducing the number of nodes.
Figure~\ref{fig:n_nodes_m} shows that in the multi-hop case, the number of nodes for Method~B is about 60\% to 70\% of that for Method~Z in the complete graph, and about 40\% in the grid graph.
In the case of single-hop, from Figure~\ref{fig:n_nodes_s}, it can be seen that in the complete graph, the number of nodes for Method~B is about 30\% to 50\% of that for Method~Z, and in the grid graph, it is about 10\% to 20\%.
Therefore, in both cases, the number of nodes can be decreased by outputting BDDs, and the effect is considered to be more pronounced for grid graphs.

\section{Conclusions}\label{s:conclusions}
In this paper, we proposed an efficient method to compute the TVN reliability.
The proposed method first constructs a ZDD representing a set of journeys, then constructs a BDD representing a family $\S_G$ of subgraphs containing $s$-$z$ journeys, and then applies dynamic programming to the obtained BDD to compute the TVN reliability.
Experimental results show that the proposed method is up to 37,372 times faster than the existing method and can significantly reduce the TVN reliability computation time.
Future work includes investigating the application of FBSJE, the main technical contribution of this research, to problems such as betweenness centrality~\cite{oettershagen_temporal_2022} and influence spread~\cite{maehara_exact_2017} in temporal graphs.
Since experiments showed that the computation time of the superset operation is a bottleneck, it is necessary to consider an efficient algorithm that avoids the superset operation and constructs a BDD that represents $\S_G$ directly from the input.



\bibliographystyle{unsrt} 
\bibliography{ref}





\end{document}